# Circuit Design Methods for Quantum Separator (QS) and Systems to Use Its Output

Paul J. Werbos, May 31, 2010

## ABSTRACT

The underlying dynamics ($\partial_t\psi=iH\psi$) of quantum electrodynamics are symmetric with respect to time (T and CPT), but traditional calculations and designs in electronics and electromagnetics impose an observer formalism or causality constraints which *assume* a gross asymmetry between forwards time and backwards time. In 2008, I published a paper in the *International Journal of Theoretical Physics* (see http://arxiv.org/abs/0801.1234) which describes how to construct physics based on the dynamics alone, without these extraneous assumptions. It pointed out that this changes certain predictions of physics in a testable way, and that evidence from experiment favors the new and simpler theory. This disclosure follows up on that paper, by describing methods for circuit design based on the new physics. It provides a striking example – how to design a quantum separator (QS), which separates out the eigenfunctions which supply ordinary time-forwards free energy from the time-inverted eigenfunctions, when the QS is connected to bidirectional power supplies now under development in several places.

## FIELD OF THE INVENTION

The present invention provides a method for separating forwards propagating energy potentials from backwards propagating energy potentials, based on the exploitation of new principles in quantum theory and emergent nanotechnology systems which provide bidirectional sources of energy potential. For initial applications and testing, the QS would be designed and used as a remote source of very high frequency DC voltage. The QS is an enabling technology for new types of quantum-based communication and computing.

## DEFINITIONS AND BACKGROUND

Traditional uses of quantum electrodynamics assumed that causal relations and flows of free energy must always propagate in the forwards time direction only. This was hard-wired into the "observer formalism." However, the Schrodinger equation for quantum electrodynamics (QED), the equation which governs the changes over time of electricity and magnetism and charged particles, is actually symmetric with respect to time. It is possible to compute the predictions of QED using the same well-established dynamics without the use of the ad hoc "observer formalism." This is called "backwards time physics" (BTP) [PJW 2008, Price 1997]. It is not yet well known outside of theoretical circles, but the logic and references in the paper in the International Journal of Theoretical Physics provide substantial logical and empirical support for this position.

According to BTP, causality *usually* goes forwards in time in our macroscopic experience because our technology is largely based on *forwards-time free energy*, inherited from the formation of the sun and the earth in past history. The energy technologies in use today are all technologies for converting positive-time free-energy into other forms of positive time free energy and to heat. The traditional interpretation of the Second Law of Thermodynamics said that this is the only possible way that we could obtain free energy.

But in the meantime, a better understanding of the Second Law has also begun to emerge, both in theory [Lloyd 1997, Second Law, Werbos 2005] and in cutting edge electronic engineering [Popovic 2004, Trew 2010, Sandia 2010a,b]. Credible device modeling calculations based on traditional QED [Trew 2010, Sandia 2010a,b] indicate that it should be possible to convert *ambient* terahertz radiation (due to heat) into electricity in a variety of devices, often pursued quietly within the general class of "thermophotovoltaics" (TPV), for which there is an important growing literature. In this disclosure, TPV which are able in theory to extract energy from ambient infrared radiation will be referred to as ATPV, ambient TPV.

But devices like ATPV are *passive*. They do not draw energy from ancient sources of forwards time free energy. They draw energy from background radiation which is generally symmetric with respect to time. Thus according to BPT, their output should provide free energy which is *symmetric* with respect to time. In other words, they provide a 50-50 mix of forwards-time free energy and backwards time free energy. If their output electrode were hooked up to an ordinary electric circuit, such as a lightbulb, one effect would send electricity into the light bulb, while the other would suck it out, in equal measure, in steady state. Thus no useful energy would be produced. In this disclosure, devices like ATPV which produce a mix of positive-time free energy and backwards time free energy will be referred to as *bidirectional power supplies*.

A *quantum separator* is defined here as a circuit or system or subsystem which can input a bidirectional power supply, and separate the power streams into separate streams which are purely or predominantly positive-time and purely or predominantly negative time, or which outputs just one of these streams to downstream loads.

## METHOD AND PRINCIPLES OF DESIGN

This disclosure will first illustrate the method with a nonautonomous linear circuit, modulated by a clock circuit. Given just one source of time-forwards external voltage V(t) (other than ground and the clock itself), a finite-node nonautonomous linear circuit is governed by:

$$V(k,t) = \int^t T_k(t,\tau) V(\tau) d\tau \qquad (1)$$

where $T_k$ is an impulse response function, where k is one of m nodes in the circuit, and where I have suppressed information about the initial conditions for simplicity. (In a stationary steady state system, we may take minus infinity as the initial time.) In the special case of an autonomous or time-invariant (LTI) circuit, not modulated by a clock, $T_k$ is a function of t-τ.
    In classical design, V(τ) is a source of time-forwards free energy, and:

$$T_k(t,\tau) < 0 \text{ for } t < \tau \qquad (2)$$

According to BTP [PJW2009], the laws governing the backwards flow of free energy are the same as those governing the forwards flow, except for time reversal. Thus for a symmetric bidirectional power source outputting a voltage V(τ):

$$V_\uparrow(\tau) = V_\downarrow(\tau) = V(\tau) \tag{3}$$

$$V_\uparrow(k,t) = \int^t T_k(t,\tau)V_\uparrow(\tau)d\tau \tag{4}$$

$$V_\downarrow(k,t) = \int_t T_k(\tau,t)V_\downarrow(\tau)d\tau \tag{5}$$

The physical voltage (actual $A_0(x,t)$) which we would observe, say, by connecting a light bulb from node k to ground, would be:

$$V_{total}(k,t) = V_\uparrow(k,t) - V_\downarrow(k,t) \tag{6}$$

Of course, this will always be zero if V(t) is constant move time and if $T_k$ is a linear autonomous system.

The voltages V here are all voltages relative to the ground connection of the bidirectional power supply and of the quantum separator, which are assumed to be the same, without loss of generality.

The key to designing a simple QS is to take advantage of a clock signal, which makes the system nonautonomous. For example, if an ATPV is connected to a simple switch, BPT does predict observeable flows for a brief time when the switch is turned on. (This is analogous to a classical capacitor, which does not allow any DC current to flow in steady state when it is hooked up to a constant DC power source like a battery, but does allow current flow immediately after a switch is turned on connecting it to the battery, as the capacitor charges up.) More precisely, if one end of the switch is connected to the main output of a TPV, and the other end is connected to a transmisison line with a delay time of one nanosecond, then backwards time free energy will flow along the line during the nanosecond *before* the switch is turned on, but ordinary forwards time free energy will not. Likewise, forwards time free energy will flow along the line during the nanosecond after the switch is turned on, but backwards time free energy will not. This is the main phenomenon which makes quantum separation possible.

The next section of this disclosure will illustrate by example a series of methods for exploiting and generalizing this approach to quantum separation in circuit design, and in the design and analysis of other circuits relying on bidirectional power supplies more a mix of forwards-flowing and backwards-flowing power supplies. The example which starts this section, using a simple analytical treatment of voltage flows, is merely one member of the larger class of designs which can be used to implement this method.

# EXAMPLE OF A QS CIRCUIT

# Figure 1. Example of a QS Circuit

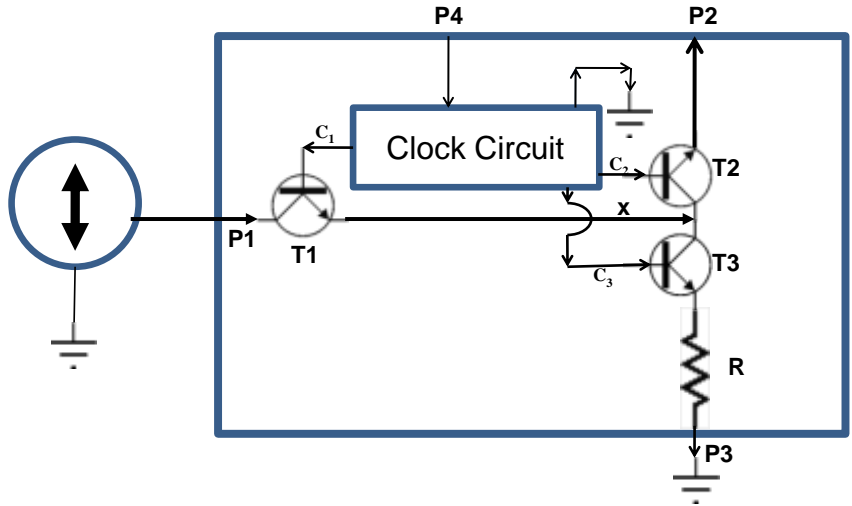

Figure 1 gives a circuit diagram for a basic QS circuit. The QS proper is the circuit enclosed in a blue rectangle. The blue circle with a two-way arrow in it symbolizes a bidirectional power supply.

P1 and P2 are the most important ports in the design. P1 takes the input from the bidirectional power supply. V(t) is just the input voltage at P1. P2 is the port which supplies ordinary DC voltage (though modulated at a high frequency) to a classical load. In the simple design here, the resistor R is scaled to match the impedance of transmission line X, so that the negative time energy is transferred to ambient cooling and to ground [Carter 1990, Hall 2000]; this ensures that users connecting loads to port P2 can treat the circuit as a classical power source. A lossless transmission line x conveys a voltage wave downstream with no change in the wave form and a fixed delay time, $t_x$ [Carter 1990, Hall 2000]. (Likewise, the characteristic impedance of a loss transmission line is a real number, sqrt(L/C); we set R=sqrt(L/C).) The three transistors – T1, T2 and T3 – may be chosen so that they may be modeled closely enough as tunable switches, turned on and off by the clock signals $c_1$, $c_2$ and $c_3$ coming from the clock circuit. The clock circuit is a classical circuit which any qualified electrical engineer could design (from many examples published in the literature), so as to produce the desired schedule of turning the transistors on and off.

A small amount of power is needed (port 4) to power the clock circuit. However, a typical transistor used in amplifiers requires only 1% as much energy as the signals they amplify; thus the power needed at port 4 would only be a small fraction of the power emitted by port P2. Furthermore, modern transistors can be chosen with delay factors as small as 10 picoseconds [IBM 2010]; thus for a reasonable approximation to predicting the behavior of this circuit, with suitable chosen transistors, we may assume that all of the delay from P1 to T2 and T3 is due to the simple line impedance and delay of x, which could be chosen for convenience to be about 1 nanosecond.

To complete the design of this simple example, it is only necessary to develop a schedule for the clock pulses. This is illustrated in Figure 2.

# Figure 2. Clock Pulses and Operation Over Time

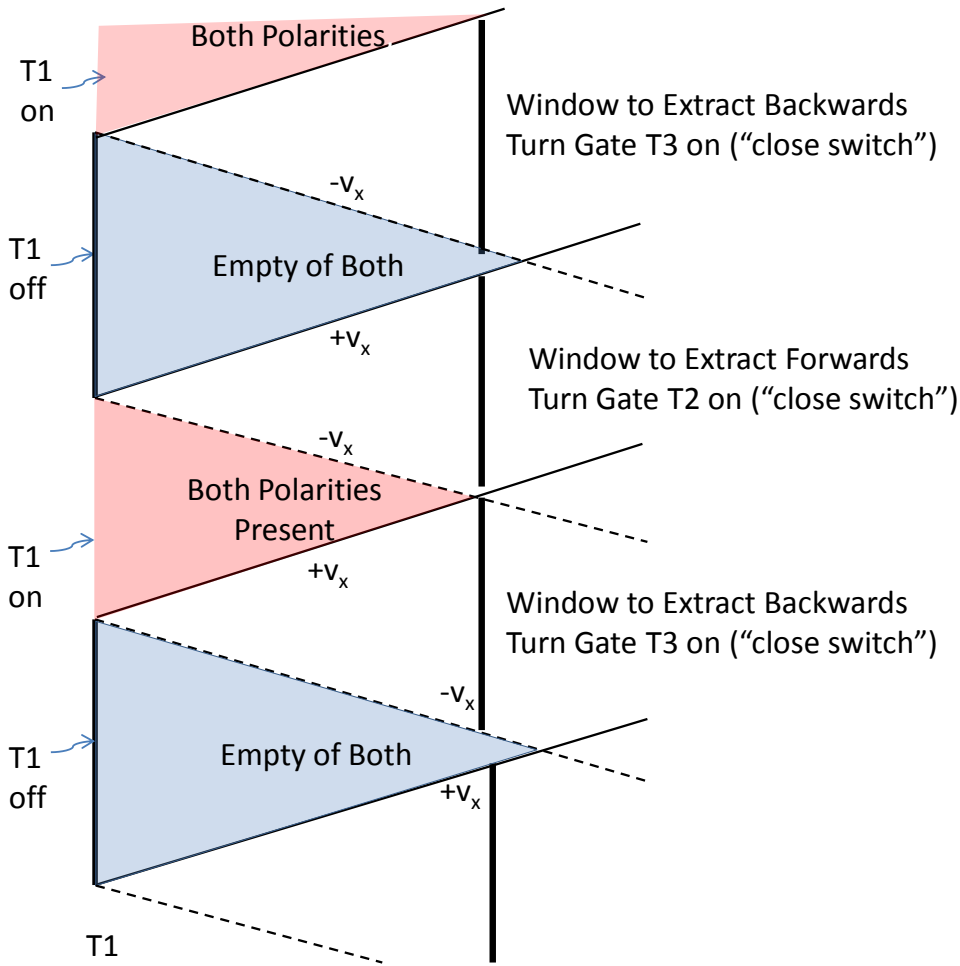

This figure shows the path of the forward propagating voltage signals (solid lines tilted up to the right) and backwards propagating voltage signals (dashed lines), across time and distance from T1 to T2 and T3. Time runs up along the y axis.

This figure assumes that the propagation of the voltage signals from T1 to T2 and T3 obeys the dynamics of a lossless transmission line [Carter 1990, Hall 2000.]. As Carter points out, the velocity of propagation, $v_x$, mainly depends on the dielectric constant of the surrounding medium. Most important is the time $t_x$ required for the signals to go from T1 to T2 or to T3. In implementing the system, the clock signals should be controlled somewhat more precisely than is shown in this figure, but it should follow the basic pattern you see here. The primary clock signal, $c_1$, should turn T1 on for a period of $2t_x$, and then off for $2t_x$, then on again, and so on; this would ensure that the propagation lines you see here cross exactly at T2 and T3. In an ideal system, $c_2$ and $c_3$ would also follow a pattern of going on for $2t_x$ and off for $2t_x$, cycling indefinitely, but the cycle for $c_2$ would start at a time $t_x$ earlier than the cycle for $c_1$, and $c_3$ would start later.

In implementing a less idealized version of this system, robust with respect to losses, one could use a package like SPICE to simulate the circuit in forwards time, to represent equation 4. One could use the same package, but with clock signals reversed in time and the predicted voltage outputs reversed in time after the run, to represent equation 5. The "on" windows for T2 and T3 could be adjusted, along with other design parameters, in order to better achieve whatever measure of performance is desired. Likewise, in implementing Figure 1, a wide variety of delay times $t_x$ could be used and evaluated, on the order of a nanosecond. For example, the transmission line x might be an on-chip interconnect with a delay time ($t_x$) of half a nanosecond. In a prototype version of this circuit, the transmission line might be an interior wire of a circuit board.

Some have claimed that an ideal ATPV should be able to generate a watt of output from a centimeter square primary collector. In an ideal system following figures 1 and 2 exactly, the QS would be outputting an average of half a watt, provided as a square wave voltage varying between two watts and zero, with a frequency of 250 megahertz if $t_x$ is one nanosecond, or 500 megahertz if it is half a nanosecond. Of course there is a large literature on how to use and filter time-varying classical power sources, such as what P2 would provide.

## GENERALIZATION AND EXTENSIONS

The methods of design and analysis described here can also be extended to the nonlinear case.

For example, many nonlinear circuits can be described as linear circuits modified by including transistors whose gate voltage comes from the circuit proper, not fixed in advance by a fixed clock. To simulate and design such a circuit, one would iterate so as to solve for a self-consistent solution for $V_\uparrow$ and $V_\downarrow$, which has the following properties. The total voltage at each node – *including* the gate voltage at each transistor across all times t – is calculated by using equation 6. That gate voltage is then held fixed, as $V_\uparrow$ is calculated in forwards time, by an ordinary circuit simulation package. It is also held fixed when when $V_\downarrow$ is calculated in backwards time, with the schedule of gate voltages reversed in time. There are many numerical approaches which can be used in this kind of iteration, such as successive overrelaxation and so on. Of course, it is possible to develop a larger integrated forwards-backwards simulation package which integrates the numerical integration and the forward and backward circuit simulations seamlessly.

Another obvious extension is to make use of the backwards time free energy available at port P3, if the resistor R and the connection to R are removed. In many cases, it would be more convenient to insert such a resistor and connection to port P2, so as to have a pure source of backwards-time free energy available at P3, for applications such as communication or computing.

To apply this methodology for circuit analysis and design, one must take special care when the ordinary time-forwards calculation for impulse response (equation 4) uses complex numbers. In some ordinary calculations, an imaginary voltage may be used to represent derivatives in forwards time (related to phase shifts in circuits). In other standard time-forwards calculations, an imaginary potential V or an imaginary self-energy term $\Gamma$ may be used to represent dissipation effects in forwards time. In running those calculations in backwards time (equation 5), one is effectively taking the complex conjugate of the original dynamics.

Dissipation still takes place in the direction opposite the direction in which free energy is flowing, allowing the use of traditional simulation packages as described here, but the corresponding Schrodinger equation expressed in forwards time would show a sign reversal for the imaginary self-energy term. In a mathematical sense, the time symmetry being exploited here is more like "CPT" – a fundamental property of quantum field theory – rather than "T."

operating in pulsed regime limited by self-heating, *Applied Physics Letters* 94, 222106 (2009); doi:10.1063/1.3147217 (*3 pages*), http://apl.aip.org/applab/v94/i22/p222106_s1?view=fulltext

[Werbos 2005] P. Werbos, Order From Chaos: A Reconsideration of Fundamental Principles, *Problems of Nonlinear Analysis in Engineering Systems*, No.3, Col. 11, 2005. http://arxiv.org/abs/cond-mat/0411384

[Werbos 2008] P. Werbos, Bell's Theorem, Many Worlds and Backwards-Time Physics: Not Just a Matter of Interpretation, *Int'l J Theoretical Physics*, April 2, 2008 (epub date), http://arxiv.org/abs/0801.1234

# CLAIMS

1. The quantum separator (QS) as defined in this disclosure.

2. The use of external clock inputs to analyze, simulate or produce a quantum separator or other systems which receive bidirectional energy inputs, or a mix of energy inputs which are partly positive-time and partly negative-time in nature.

3. The use of equations 5 or 6 in the design or operation of electronic or electro-optic or photonic devices.

4. The use of a consistent solution of backwards causal simulation, combined with forwards causal simulation, in order to analyze, predict or design systems or devices which have at least some input of backwards-time free energy as part of their principle of operation.

5. Electronic or photonic or electro-optic systems or devices which have at least some input of backwards-time free energy as part of their principle of operation.

# PRIOR WORK AIMED AT SIMILAR INVENTIONS

This invention is relatively unique. The references above are the most closely related prior sources.

As cited – there has been considerable work on TPVs and passive (bidirectional) power sources. But this disclosure does not claim an invention in that area. It claims a method for actually making use of such of such power sources. Information Property protection is an essential first step before making the new system available for use by people and organizations with such sources.

Huw Price (cited above) did propose a kind of thought experiment, involving the use of tempearture measurements on telescopes to try to to observe sources of radiation flowing backwards through time from other parts of the universe (where the boundary conditions in time

May be different from what they are in our vicinity). But there was never any attempt to reduce this idea to an actual device, nor was there discussion of backwards-time free energy as such.

Many, many less informed people have proposed the development of systems for communicating information faster than the speed of light or backwards through time by exploiting the "nonlocal" effects present in Bell's Theorem experiments. But J.S. Bell, in his classic book *The Speakable and Unspeakable in Quantum Mechanics,* has explained why that is impossible. Prof. YanHua Shih et al, in the first issue of Physical Review Letters in 2000, reported a quantum delay eraser experiment which suggested a more sophisticated way to try to communicate backwards through time; however, the underlying mathematics are essentially the same as those of Bell's Theorem experiments. They convey evidence that backwards causal effects are happening at a microscopic level, but do not give any idea about how they could be used in macroscopic devices. None of these systems, using forwards-time free energy as the only way to power their experiments, have been able to achieve the goals of real communications or computing backwards in time.

Advanced and retarded Green's functions, which move mathematically forwards and backwards in time, are used very often as a mathematical device in physics. However, the use of these functions is generally constrained by "causality constraints" or passivity constraints which do not allow the kind of technologies described here.

As Price pointed out in his classic article, even the best researchers often find it difficult to think objectively about physical phenomena in which temporal effects are truly symmetric – even though Einstein's special relativity clearly impels us to learn a new way of thinking. Many of the contents on this invention, like special relativity itself, should seem obvious *later, in retrospect, to many engineers* – once the new way of thinking is assimilated. The invention itself may play a crucial role in making that assimilation possible.

## FURTHER EXPLANATION OF EQUATIONS 3 TO 6

This disclosure has already described *how to use* equations 3 through 6 in designing circuits which get their power from a constant bidirectional source, or, more generally, from any mix of time-forwards and time-backwards sources of voltage.

This section will say more about the equations themselves.

First, these equations are fully consistent with the extensive experience the world already has in circuit design. So far, that experience has all involved circuits which make use of forwards-time free energy only. In that special case, $V_\downarrow=0$ everywhere, equation 3 does not apply, and equations 4 through 6 reduce to equation 1, the usual situation in the past.

Equation 5 is just the reflection through time of equation 4. Symmetry through time is a fundamental property of the "Schrodinger equation" of quantum electrodynamics, and a key feature of BTP (Werbos 2008).

Equation 6 simply expresses the concrete realities of situations like the lightbulb example described earlier in this disclosure. For a more formal treatment – in a bidirectional power supply, "floating in space" without using a source of ordinary time-forwards free energy, the flows of energy of the entire system are a byproduct of the statistics of the quantum grand ensemble Boltzman distribution. That distribution gives a certain probability to each "forwards" many-body eigenfunction which would move current transport forwards from the power supply to the transmission line x in our sample circuit. But for every such eigenfunction, there is a time-reversed version of that same eigenfunction, which would carry charge in the reverse direction, with the same probability (since the Boltzmann distribution is also time-symmetric). What we have called backwards-flowing free energy is really just the action of these eigenfunctions.

This analysis suggests a further question, and a further family of possible technologies. Is it possible to create a quantum entanglement between the forwards eigenfunctions and the backwards eigenfunctions, leading to some kind of entanglement across time, between past and future? Can such an entanglement be used to support a new type of quantum computing? Empirical tests will be needed to evaluate these possibilities. Those empirical tests will need to start with a reliable source of both types of eigenfunction. The invention disclosed here fulfills that requirement.